\documentclass[11pt,a4paper]{article}
\usepackage[english]{babel}
\usepackage{amsmath,amsthm,amssymb,epsfig,latexsym,color}
\usepackage[numbers,square,sort&compress]{natbib}
\usepackage{color}

\newcommand{\so}{\scriptscriptstyle \rm I}
\newcommand{\st}{\scriptscriptstyle \rm I\hspace{-1pt}I}

\newcommand{\be}[1]{\begin{equation}\label{#1}}
\newcommand{\ee}{\end{equation}}
\newcommand{\bu}{\bar u}
\newcommand{\bv}{\bar v}

\newcommand{\bet}{\bar \eta}

\newcommand{\bxi}{\bar\xi}

\newcommand{\tr}{\mathop{\rm tr}}
\newcommand{\diag}{\mathop{\rm diag}}

\newcommand{\kb}{\tilde \kappa}

\newcommand{\kp}{ \kappa^+}
\newcommand{\km}{ \kappa^-}

\newcommand{\ben}{\begin{eqnarray}}
\newcommand{\een}{\end{eqnarray}}

\newtheorem{prop}{Proposition}[section]
\newtheorem{cor}{Corollary}[section]
\newtheorem{Def}{Definition}[section]
 \makeatletter
 \@addtoreset{equation}{section}
 \makeatother

\begin{document}

\vspace{4pt}

\begin{center}
\begin{LARGE}
{\bf Overlap between usual and modified Bethe vectors}
\end{LARGE}

\vspace{44pt}

\begin{large}
{S. Belliard${}^\dagger$ and N.~A.~Slavnov${}^\ddagger$\   \footnote{samuel.belliard@univ-tours.fr, nslavnov@mi-ras.ru}}
\end{large}

 \vspace{6mm}

  \vspace{4mm}

${}^\dagger$ {\it Institut Denis-Poisson, Universit\'e de Tours, Universit\'e d'Orl\'eans, Parc de Grammont, 37200 Tours, FRANCE}

 \vspace{4mm}

${}^\ddagger$  {\it Steklov Mathematical Institute of Russian Academy of Sciences,\\ 8 Gubkina str., Moscow, 119991,  RUSSIA}

\end{center}

\vspace{10mm}

\begin{abstract}
We consider the overlap of Bethe vectors of the XXX spin chain with a diagonal twist  and the modified Bethe vectors with a general twist. We find a determinant representation for this overlap under one additional condition on the twist parameters. Such objects arise in the calculations of nonequilibrium physics.
\end{abstract}

\vspace{10mm}


\section{Introduction}

Integrable quantum spin chains are a powerful tool for studying nonequilibrium physics. They have parameters that can be considered time-dependent. A simple example is an external magnetic field given by a Heaviside function that turns the field on or off. One can also consider the change in the parameters that define the boundary conditions. For example, a chain of $N$ spins can be closed periodically or in the form of the M\"obius strip. In the general case, the classification of integrable boundary parameters follows from the Quantum Inverse Scattering Method and related quantum algebras. In particular, the twisting of isotropic closed spin chains is given by an arbitrary invertible matrix.

To study the behavior of quantum integrable models when the boundary conditions change, we should solve the eigenproblem for any integrable boundary parameters and calculate the overlap between the states corresponding to different values of these parameters. For this, we use the Quantum Inverse Scattering Method \cite{FadST79, FadLH96,BogIK93L}. In this method, the eigenvectors of quantum Hamiltonians are constructed using the algebraic Bethe ansatz, and the roots of the Bethe equations parameterize the spectrum. A modified version of the Bethe ansatz is needed in the case of so-called "off-diagonal boundaries" \cite{BelC13}. The spectrum of such models is given by modified Bethe equations, which contain an inhomogeneous term \cite{CYSW13a,CYSW13c}. In this case, there is an additional restriction on the number of Bethe equations.

In a recent series of papers, we studied the scalar products of modified Bethe vectors \cite{BelSV18sc,BelSV18,BelS19}. We found that the well-known formulas of the standard algebraic Bethe ansatz smoothly transform into a modified form that has the same structure. This applies both to scalar products of vectors of a general form \cite{Kor82} and to scalar products containing the eigenvectors of a modified transfer matrix \cite{Gau72,Gaud83,Kor82,Sla89}. In this paper, we consider the overlap (scalar product) of two vectors corresponding to different transfer matrices. One of the vectors corresponds to a transfer matrix with a diagonal twist. It can be constructed within the framework of the standard algebraic Bethe ansatz. The second vector is a modified Bethe vector, which corresponds to a transfer matrix with a twist of the general form. To construct this vector, one needs to use the modified Bethe ansatz. We show that when one additional condition is imposed on the twist parameters, such an overlap has a  compact determinant representation. This opens a possibility to study quenches in the XXX chains of an arbitrary spin.

The paper is organized as follows. In section~\ref{S-QISM}, we recall basic notions of the Quantum Inverse Scattering Method. Here we also introduce a special notation used below. In section~\ref{S-BV}, we introduce a modified transfer matrix and construct modified Bethe vectors. Section~\ref{S-MDID} is devoted to a modified Izergin determinant \cite{BelSV18sc}. Here we derive a new representation which contains a set of arbitrary complex numbers. In section~\ref{S2}, we consider the overlap between Bethe vectors with diagonal twist and modified Bethe vectors of general type. We show that under one additional constraint between the twist parameters, this overlap has a determinant representation. Appendix~\ref{A-PMID} contains some auxiliary formulas for the modified Izergin determinant  and proofs of some propositions used in the calculation.


\section{Quantum inverse scattering method \label{S-QISM}}

To formulate the  problem within the Quantum Inverse Scattering Method framework, we consider a monodromy matrix
\be{MonoT}
T(u)=\begin{pmatrix} t_{11}(u)& t_{12}(u)\\ t_{21}(u)&t_{22}(u)\end{pmatrix}.
\ee
The matrix elements $t_{kl}(u)$ act in a Hilbert space $\mathcal{H}$ and depend on a complex parameter $u$. The commutation relations between $t_{kl}(u)$
are defined by an RTT-relation
\be{RTT}
R_{12}(u-v)T_1(u)T_2(v)=T_2(v)T_1(u)R_{12}(u-v).
\ee
Here $T_1(u)=T(u)\otimes \mathbf{1}$ and $T_2(u)=\mathbf{1}\otimes T(u)$, where $\mathbf{1}$ is the identity matrix in $\mathbb{C}^2$.
The $R$-matrix $R_{12}(u)$ acts in $\mathbb{C}^2\otimes\mathbb{C}^2$ and has the form
\be{Rmat}
R(u)=\frac uc (\mathbf{1}\otimes\mathbf{1})+P.
\ee
Here $P$ the permutation operator such that $Px\otimes y=y\otimes x$ for any $x,y\in \mathbb{C}^2$, and $c$ is a constant.
The $R$-matrix \eqref{Rmat} has the following property:
\be{RKK}
[R_{12}(u-v),K_1K_2]=0,
\ee
for any $2\times 2$ matrix $K$.
This property leads to the fact that a matrix $KT(u)$ satisfies the RTT-relation \eqref{RTT}.

The trace of the monodromy matrix $\tr T(u)=t_{11}(u)+t_{22}(u)$ is called a transfer matrix. Due to the RTT-relation \eqref{RTT},
\be{commTr}
[\tr T(u),\tr T(v)]=0
\ee
for any complex $u$ and $v$. This property of the transfer matrix allows us to consider it as a generating function of the integrals of motion of a quantum model. Due to the cyclicity of the trace, this model has periodic boundary conditions. The trace of the twisted monodromy matrix $\tr(KT(u))$ can also be used to generate integrals of motion. However, they satisfy boundary conditions of a more general form.

Let us define a highest weight representation by $V(\lambda_1(u), \lambda_2(u))$, where $\lambda_i(u)$
are some complex valued functions, and the highest weight vector $|0\rangle\in\mathcal{H}$ is defined by
\ben\label{HWRG}
t_{ii}(u)|0\rangle=\lambda_i(u)|0\rangle, \qquad t_{21}(u)|0\rangle=0.
\een
The action of the operator $t_{12}(u)$ on $|0\rangle$
is free. A state obtained by the successive action of the $t_{12}$ operators on the highest weight vector is called a Bethe vector
\ben\label{BVdef}
 |\Phi(\bu)\rangle=  t_{12}(u_1)\dots t_{12}(u_n)|0\rangle.
\een
Here $n=0,1,\dots$, and  $\bu=\{u_1,\dots,u_n\}$.  If the parameters $\bu$ are generic complex numbers, then $|\Phi(\bu)\rangle$ is called the off-shell Bethe vector. Under certain constraint on the parameters $\bu$ (see section~\ref{S-BV}), the vector \eqref{BVdef} becomes an eigenvector
of the transfer matrix. Then it is called the on-shell Bethe vector.

To study scalar products we also introduce a dual vector $\langle 0|\in\mathcal{H}^*$ defined by
\ben\label{dHWRG}
\langle 0|t_{ii}(u)=\lambda_i(u)\langle 0|, \qquad \langle 0|t_{12}(u)=0,
\een
and a condition $\langle 0|0\rangle=1$. The functions $\lambda_i(u)$ are the same as in \eqref{HWRG}. Dual Bethe vectors are constructed by successive action of the $t_{21}$ operators on $\langle0|$:
\be{dBVdef}
\langle \Phi(\bu)|=\langle 0|t_{21}(u_1)\dots t_{21}(u_n).
\ee

\subsection{Notation}\label{SS-NOT}

To simplify formulas, we introduce three rational functions
\ben\label{gfh}
g(u,v)=\frac{c}{u-v},\quad f(u,v)=1+g(u,v)=\frac{u-v+c}{u-v}, \quad h(u,v)= \frac{f(u,v)}{g(u,v)}= \frac{u-v+c}{c},
\een
where $c$ is the constant entering the $R$-matrix.
It is easy to see that the functions \eqref{gfh} have the following  properties:
\ben\label{x-gfh}
\chi(u,v)\Bigr|_{c\to -c}=\chi(v,u), \qquad \chi(-u,-v)=\chi(v,u),\qquad  \chi(u-c,v)=\chi(u,v+c),
\een
where $\chi$ is any of the three functions. Besides, we have
\ben\label{gfh-prop}
g(u,v-c)=\frac{1}{h(u,v)},\quad h(u,v+c)= \frac{1}{g(u,v)}, \quad f(u,v+c)=\frac{1}{f(v,u)}.
\een

Below we deal with sets of complex parameters. We denote them by a bar, for example, $\bu=\{u_1,\dots,u_n\}$. Notation
$\bu\pm c$ means that $\pm c$ is added to all the arguments of the set $\bu$.
The notation $\bu_k$ refers to the  subset  $\bu_k=\bu\setminus u_k$.

To make the formulas more compact, we use a shorthand notation for the products of the rational functions \eqref{gfh}, the operators
$t_{kl}(u)$ \eqref{MonoT}, and their vacuum eigenvalues $\lambda_i(u)$ \eqref{HWRG}. Namely, if  the  function (operator) depends on a set of variables, then this means the product over the corresponding set. For example, if $\bu=\{u_1,\dots,u_n\}$, then
\be{shn}
t_{kl}(\bu)=\prod_{j=1}^n t_{kl}(u_j), \quad \lambda_{i}(\bu)=\prod_{j=1}^n  \lambda_{i}(u_j), \quad f(z,\bu)=\prod_{j=1}^n f(z,u_j), \quad f(\bu_k,u_k)=\prod_{\substack{j=1\\j\ne k}}^nf(u_j,u_k),
\ee
and so on. Note that the RTT-relation \eqref{RTT} implies $[t_{kl}(u),t_{kl}(v)]=0$. Thus, the first product in \eqref{shn} is well defined. Later we will extend this convention to the products of matrix elements of the twisted monodromy matrix.

Notation $f(\bu,\bv)$ means   the  double product over the sets $\bu$ and $\bv$.
By definition, any product over the empty set is equal to $1$. A double product is equal to $1$ if at least one of the sets is empty.

Finally, for any set of complex parameters $\bu=\{u_1,\dots,u_n\}$ such that $n\ge 2$, we also introduce special products of the $g$-functions
\be{DD-def}
\Delta(\bu)=\prod_{1\le k<j\le n}g(u_j,u_k), \qquad \Delta'(\bu)=\prod_{1\le k<j\le n}g(u_k,u_j).
\ee
It is easy to see that $\Delta(\bu)=(-1)^{n(n-1)/2}\Delta'(\bu)$. For $n=0,1$, we set $\Delta(\bu)=\Delta'(\bu)=1$ by definition.


\section{Twisted monodromy matrix\label{S-BV}}

We consider two twisted monodromy matrices $K^{(\ell)}T(u)$, $\ell=1,2$.
The first one corresponds to the twist matrix $K^{(1)}=\diag(1,\alpha)$, where $\alpha$ is a complex parameter. Then the twisted transfer matrix has the form
\be{ttrans}
t^{(1)}(u)=t_{11}(u)+\alpha t_{22}(u).
\ee
We call it a diagonal transfer matrix. It generates a model with quasi-periodic boundary conditions. The eigenvectors (resp. dual eigenvectors) of the diagonal transfer matrix have the form \eqref{BVdef} (resp. \eqref{dBVdef}) %
\ben\label{BVdef1}
 |\Phi^{(1)}(\bu)\rangle=  t_{12}(\bu)|0\rangle, \qquad \langle \Phi^{(1)}(\bu)|=\langle 0|t_{21}(\bu).
\een
We have added an extra superscript to the vectors to emphasize that they refer to the transfer matrix $t^{(1)}(u)$.

If the parameters $\bu=\{u_1,\dots,u_n\}$ are arbitrary complex numbers, then the vectors \eqref{BVdef1} are off-shell.  If these parameters satisfy twisted Bethe equations

\be{BEt}
\lambda_1(u_j)f(\bu_j,u_j)=\alpha\lambda_2(u_j)f(u_j,\bu_j),\qquad j=1,\dots,n, 
\ee
then
\ben
t^{(1)}(v)| \Phi^{(1)}(\bu)\rangle=\Lambda^{(1)}(v|\bu)| \Phi^{(1)}(\bu)\rangle , \qquad \langle \Phi^{(1)}(\bu)|t^{(1)}(v)=\Lambda^{(1)}(v|\bu)\langle \Phi^{(1)}(\bu)|,
\een
where
\be{EigenV}
\Lambda^{(1)}(v|\bu)=\lambda_1(v)f(\bu,v)+\alpha \lambda_2(v)f(v,\bu).
\ee
The vectors $| \Phi^{(1)}(\bu)\rangle$ and $\langle \Phi^{(1)}(\bu)|$ are called on-shell provided the conditions \eqref{BEt} are fulfilled.
In the case of the XXX spin-$1/2$ chain, the number of parameters $\{u_1,\dots,u_n\}$ does not exceed the length of the chain.

The second deformation corresponds to the twist matrix of the general form
\ben\label{gentwist}
K^{(2)}=\begin{pmatrix} \tilde \kappa & \;\kappa^+\\ \;\kappa^-&\kappa \end{pmatrix}, 
\een
where the entries of $K^{(2)}$ are arbitrary complex numbers.
This twist produces models with non-diagonal boundary conditions. The standard procedure of the algebraic Bethe ansatz should be modified in this case \cite{BelP15,BelSV18}. We present the twist matrix in the form $K^{(2)}=BDA$ where
\be{Mat-Tf}
A=\sqrt\mu\begin{pmatrix} 1&\frac{\rho_2}{\km}\\ \frac{\rho_1}{\kp}&1\end{pmatrix},\qquad B=\sqrt\mu\begin{pmatrix} 1&\frac{\rho_1}{\km}\\ \frac{\rho_2}{\kp}&1\end{pmatrix},
\qquad D=\begin{pmatrix} \tilde \kappa-\rho_1 &0\\ 0&\kappa-\rho_2 \end{pmatrix}.
\ee
Here
\be{mu-Tf}
\mu=\frac{1}{1-\frac{\rho_1\rho_2}{\kp\km}},
\ee
and the parameters $\rho_\ell$ satisfy a condition
\ben
\rho_1\rho_2-(\kappa \rho_1+\tilde \kappa \rho_2)+ \kappa^+ \kappa^-=0.
\een
Then a twisted transfer matrix is defined as follows:
\be{ttrans1}
t^{(2)}(u)=\tr(K^{(2)}T(u))=\tr(D \overline{T}(u)).
\ee
Here $\overline{T}(u)$ is a modified monodromy matrix
\ben\label{MbarT}
\overline{T}(u)=AT(u)B=\begin{pmatrix} \nu_{11}(u)& \nu_{12}(u)\\ \nu_{21}(u)&\nu_{22}(u)\end{pmatrix}.
\een
The entries of $\overline{T}(u)$ have the following expressions in terms of the initial monodromy matrix elements:
\ben\label{nutot11}
\nu_{11}(z)=\mu \Big(t_{11}(z)+\frac{\rho_2}{\kp}t_{12}(z)+\frac{\rho_2}{\km}t_{21}(z)+\frac{\rho_2^2}{\km\kp}t_{22}(z)\Big),\\
\label{nutot22}
\nu_{22}(z)=\mu \Big(t_{22}(z)+\frac{\rho_1}{\kp}t_{12}(z)+\frac{\rho_1}{\km}t_{21}(z)+\frac{\rho_1^2}{\km\kp}t_{11}(z)\Big),\\
\label{nutot12}
\nu_{12}(z)=\mu \Big(t_{12}(z)+\frac{\rho_1}{\km}t_{11}(z)+\frac{\rho_2}{\km}t_{22}(z)+\frac{\rho_1\rho_2}{(\km)^2}t_{21}(z)\Big),\\
\label{nutot21}
\nu_{21}(z)=\mu \Big(t_{21}(z)+\frac{\rho_1}{\kp}t_{11}(z)+\frac{\rho_2}{\kp}t_{22}(z)+\frac{\rho_1\rho_2}{(\kp)^2}t_{12}(z)\Big).
\een

The modified Bethe vectors (resp. the dual modified Bethe vectors) related to the twisted transfer matrix \eqref{ttrans1} can be constructed from the new entries $\nu_{12}(z)$ (resp. $\nu_{21}(z)$). They are given by
\ben\label{Phinu}
| \Phi^{(2)}(\bu)\rangle=\nu_{12}(\bu)|0\rangle, \qquad \langle \Phi^{(2)}(\bu)|=\langle 0|\nu_{21}(\bu).
\een
Here we have used the convention on the shorthand notation for the products of the operators $\nu_{ij}(u)$.

If  $\bu=\{u_1,\dots,u_N\}$ is a set of arbitrary complex numbers,  then vectors \eqref{Phinu} are off-shell. However, if the set $\bu$
satisfies inhomogeneous Bethe equations
\ben\label{MBE}
(\tilde\kappa-\rho_1)\lambda_1(u_j)f(\bu_j,u_j)=(\kappa-\rho_2)\lambda_2(u_j)f(u_j,\bu_j)+(\rho_1+\rho_2)\lambda_1(u_j)\lambda_2(u_j)g(u_j,\bu_j),
\een
for $j=1,\dots,N$, then they are eigenvectors of the  twisted transfer matrix \eqref{ttrans1}
\ben
t^{(2)}(v)| \Phi^{(2)}(\bu)\rangle=\Lambda^{(2)}(v|\bu)| \Phi^{(2)}(\bu)\rangle, \qquad \langle \Phi^{(2)}(\bu)|t^{(2)}(v)=\Lambda^{(2)}(v|\bu)\langle \Phi^{(2)}(\bu)|,
\een
with an eigenvalue
\ben
\Lambda^{(2)}(v|\bu)=(\tilde\kappa-\rho_1)\lambda_1(v)f(\bu,v)+(\kappa-\rho_2)\lambda_2(v)f(v,\bu)+(\rho_1+\rho_2)\lambda_1(v)\lambda_2(v)g(v,\bu).
\een
In this case, we call the vectors $| \Phi^{(2)}(\bu)\rangle$ and $\langle \Phi^{(2)}(\bu)|$ modified on-shell vectors.

The number of parameters $\{u_1,\dots,u_N\}$ depends on the chain length and spin. In particular, in the XXX spin-$1/2$ chain, $N$ coincides with the number of sites. In the case of higher spins, $N$ exceeds the chain length (see \cite{BelSV18} for more details).

The main goal of this paper is to calculate an overlap
\ben\label{SP-def0}
S^{n,N}(\bv,\bu)=\langle \Phi^{(1)}(\bv)|\Phi^{(2)}(\bu)\rangle.
\een
Here $\langle \Phi^{(1)}(\bv)|$ and $|\Phi^{(2)}(\bu)\rangle$ are the eigenvectors of the corresponding transfer matrices,  $n=\#\bv$, and $N=\#\bu$.
Then according to Fermi's Golden rule for the transition between the two twists, we obtain
\ben
\Gamma_{1\to 2}=\frac{2\pi}{\hbar}\Big|\langle \Phi^{(1)}(\bv)|(H^{(1)}-H^{(2)})| \Phi^{(2)}(\bu)\rangle\Big|^2{\rho^{(2)}}(\bu).
\een
Here $H^{(1)}$ and $H^{(2)}$ respectively are the Hamiltonians corresponding to the transfer matrices $t^{(1)}$ and $t^{(2)}$, and $\rho^{(2)}(\bu)$ is a density of the resulting states.

In the case of the XXX spin-$1/2$ chains, we can express the Hamiltonians in terms of the twisted transfer matrices using
\begin{equation}\label{H-tau}
H^{(1)}-H^{(2)}=2 c \frac{d}{dz} \big(\log t^{(1)}(z)-\log t^{(2)}(z)\big)\Bigr|_{z=0}.
\end{equation}
Thus, we find
\ben
\Gamma_{1\to 2}=\frac{2\pi}{\hbar}\left|2 c \frac{d}{dz} \log\left(\frac{\Lambda^{(1)}(z|\bv)}{\Lambda^{(2)}(z|\bu)}\right)\right|^2_{z=0}\big|S^{n,N}(\bv,\bu)\big|^2{\rho}^{(2)}(\bu).
\een


\section{Modified Izergin determinant  \label{S-MDID}}

In various formulas for the scalar products, a modified Izergin determinant (MID) arises.

\begin{Def}
Let $\bu=\{u_1,\dots,u_n\}$,  $\bv=\{v_1,\dots,v_m\}$, and $z$ be arbitrary complex numbers.
Then  the {\rm MID} $K_{n,m}^{(z)}(\bu|\bv)$
is defined by
\be{defKdef1}
K_{n,m}^{(z)}(\bu|\bv)=\det_m\left(-z\delta_{jk}+\frac{f(\bu,v_j)f(v_j,\bv_j)}{h(v_j,v_k)}\right).
\ee
Alternatively the {\rm MID} can be presented as
\be{defKdef2}
K_{n,m}^{(z)}(\bu|\bv)=(1-z)^{m-n}\det_n\left(\delta_{jk}f(u_j,\bv)-z\frac{f(u_j,\bu_j)}{h(u_j,u_k)}\right).
\ee
\end{Def}

The proof of the equivalence of representations \eqref{defKdef1} and \eqref{defKdef2} can be found in proposition~4.1 of \cite{GorZZ14}.

For $m=n$ and $z=1$, the MID turns into the usual Izergin determinant \cite{Ize87}. The latter is equal to the partition function of the six-vertex model with domain wall boundary condition \cite{Kor82,Ize87}.

We will also use a conjugated MID
\be{CdefKdef1}
\overline{K}_{n,m}^{(z)}(\bu|\bv)=K_{n,m}^{(z)}(\bu|\bv)\Bigr|_{c\to -c}
=\det_m\left(-z\delta_{jk}+\frac{f(v_j,\bu)f(\bv_j,v_j)}{h(v_k,v_j)}\right).
\ee
Equivalently, it can be defined as follows:
\be{CdefKdef2}
\overline{K}_{n,m}^{(z)}(\bu|\bv)=(1-z)^{m-n}\det_n\left(\delta_{jk}f(\bv,u_j)-z\frac{f(\bu_j,u_j)}{h(u_k,u_j)}\right).
\ee

There exist also representations for MID, which contain additional parameters.


\begin{prop}
Let $\bu=\{u_1,\dots,u_n\}$, $\bet=\{\eta_1,\dots,\eta_n\}$, $\bv=\{v_1,\dots,v_m\}$, and $z$ be arbitrary complex numbers. Then
\be{Kmodeat01}
K_{n,m}^{(z)}(\bu|\bv)=(1-z)^{m-n}\Delta'(\bu)\Delta(\bar\eta)\det_n\left(\frac{f(u_j,\bv)}{g(u_j,\bar\eta_k)}-zh(u_j,\bar\eta_k)\right).
\ee
\end{prop}
\textsl{Proof.} Let $\bu=\{u_1,\dots,u_n\}$ be pair-wise distinct complex numbers. Consider  an $n\times n$ matrix $W$ with the elements
\be{Wjk}
W_{jk}=\frac{g(u_j,\bu_j)}{g(u_j,\bar\eta_k)}.
\ee
Here $\bet=\{\eta_1,\dots,\eta_n\}$ are arbitrary pair-wise distinct complex numbers.  The entries $W_{jk}$ are proportional to the Cauchy matrix $g(u_j,\eta_k)$. Thus,
\be{detW}
\det_n W=\frac{\Delta'(\bu)\Delta(\bu)}{g(\bu,\bar\eta)}\det_n g(u_j,\eta_k)=\frac{\Delta(\bu)}{\Delta(\bar\eta)}.
\ee
Thus, the determinant of $W$ exists and is non-vanishing.
Let us transform representation \eqref{defKdef2} as follows:
\be{defKdefv-tr}
K_{n,m}^{(z)}(\bu|\bv)=\frac{(1-z)^{m-n}}{ \det_n W}\det_n\left(f(u_j,\bv)W_{jk}-z\sum_{l=1}^n\frac{f(u_j,\bu_j)}{h(u_j,u_l)}W_{lk}\right).
\ee
The sum over $l$ is easily computable. Indeed, let
\be{Gdef1}
 G_{jk}=\sum_{\ell=1}^n\frac{W_{lk}}{h(u_j,u_l)}=\sum_{\ell=1}^n\frac{g(u_\ell,\eta_k)}{h(u_j,u_\ell)}\frac{g(u_\ell,\bu_\ell)}{g(u_\ell,\bet)}.
\ee
Then we have
\be{contint01}
\frac1{2\pi ic}\oint_{|z|=R\to\infty} \frac{g(z,\eta_k)}{h(u_j,z)}\frac{g(z,\bu)}{g(z,\bet)}\,dz=0=G_{jk}
-\frac{h(u_j,\bet_k)}{h(u_j,\bu)}.
\ee
Substituting this into \eqref{defKdefv-tr} we obtain
\be{defKdefv-tr1}
K_{n,m}^{(z)}(\bu|\bv)=\frac{(1-z)^{m-n}\Delta(\bar\eta)}{\Delta(\bu)}\det_n\left(f(u_j,\bv)\frac{g(u_j,\bu_j)}{g(u_j,\bar\eta_k)}
-z g(u_j,\bu_j)h(u_j,\bar\eta_k)\right).
\ee
Extracting the products $g(u_j,\bu_j)$ we arrive at \eqref{Kmodeat01}. It remains to note that the limits $u_j=u_k$ and $\eta_j=\eta_k$ ($j,k=1,\dots,n$) are well defined. Indeed, in this case, the prefactor $\Delta'(\bu)\Delta(\bar\eta)$ has a pole, while the determinant vanishes. Therefore, representation \eqref{Kmodeat01} is valid for any complex $\bu$ and $\bet$. \qed

Similarly, one can prove the following representation for the MID:

\be{Kmodeat00}
K_{n,m}^{(z)}(\bu|\bv)=\Delta'(\bu)\Delta(\bar\eta)\det_m\left(f(\bu,v_j)h(u_j,\bar\eta_k)-\frac{z}{g(u_j,\bar\eta_k)}\right).
\ee
Here $\bet=\{\eta_1,\dots,\eta_m\}$ is a set of arbitrary complex numbers.

\begin{cor}\label{cor-ize}
Let $\bu=\{u_1,\dots,u_n\}$, $\bet=\{\eta_1,\dots,\eta_n\}$,  and $z$ be arbitrary complex numbers. Then
\be{Kmodeat1}
\Delta'(\bu)\Delta(\bar\eta)\det_n\left(\frac{1}{g(u_j,\bar\eta_k)}-zh(u_j,\bar\eta_k)\right)=(1-z)^{n}.
\ee
\end{cor}

\textsl{Proof.} Setting $\bv=\emptyset$ in \eqref{Kmodeat01} we arrive at
\be{Kmodeat2}
\Delta'(\bu)\Delta(\bar\eta)\det_n\left(\frac{1}{g(u_j,\bar\eta_k)}-zh(u_j,\bar\eta_k)\right)=(1-z)^{n}K_{n,0}^{(z)}(\bu|\emptyset).
\ee
Then \eqref{Kmodeat1} follows from \eqref{defKdef1}.

Some other properties of the MID are given in appendix~\ref{A-PMID} (see also \cite{BelSV18sc} for a more detailed description).


\section{Overlap}\label{S2}

Let us now proceed to the calculation of the overlap \eqref{SP-def0}
\ben\label{SP-ddef}
S^{n,N}(\bv,\bu)=\langle 0|t_{21}(\bv)\nu_{12}(\bu)|0\rangle.
\een
Recall that here $n=\#\bv$ and $N=\#\bu$. Ultimately, we are interested in the case when the parameters $\bv$ and $\bu$ respectively satisfy equations \eqref{BEt} and \eqref{MBE}. However, in this paper, we consider a more general case when the parameters $\bu$ are arbitrary complex numbers. In particular, they can satisfy the system \eqref{MBE}. Moreover, at the first stage of calculations, we do not impose any restrictions on the parameters $\bv$. We only require that $n\le N$,  because otherwise, the scalar product \eqref{SP-ddef} vanishes.

\subsection{Scalar product of off-shell vectors}

It follows from \eqref{nutot12} that the vector $\mu^{-N}|\Phi^{(2)}(\bu)\rangle$  does not
depend on $\kp$.  Therefore, the scalar product $\mu^{-N}S^{n,N}(\bv,\bu)$ also does not
depend on $\kp$. On the other hand, it follows from \eqref{nutot21} that
\be{limnu21}
t_{21}(v)=\lim_{\kp\to\infty}\nu_{21}(v).
\ee
Thus,
\ben
\langle 0|t_{21}(\bv)\nu_{12}(\bu)|0\rangle=\mu^N\lim_{\kappa^+\to \infty}\langle 0|\nu_{21}(\bv)\nu_{12}(\bu)|0\rangle,
\een
where we used $\mu\to 1$ as $\kappa^+\to \infty$.
In its turn, a formula for the scalar product $\langle 0|\nu_{21}(\bv)\nu_{12}(\bu)|0\rangle$ was derived in \cite{BelSV18sc}:
\begin{multline}\label{SP-prev}
\langle 0|\nu_{21}(\bv)\nu_{12}(\bu)|0\rangle=\mu^{N+n}\left(\frac{\rho_1}{\kappa^-}\right)^{N-n}\sum_{\substack{\{\bu_{\so},\bu_{\st}\}\vdash\bu\\ \{\bv_{\so},\bv_{\st}\}\vdash\bv }}
\left(\frac{\rho_1}{\rho_2}\right)^{n_{\st}-N_{\st}}
\lambda_2(\bv_{\so})\lambda_2(\bu_{\st})\lambda_1(\bv_{\st})\lambda_1(\bu_{\so})
\\
\times f(\bv_{\so},\bv_{\st})f(\bu_{\st},\bu_{\so})
K^{(1/\mu)}_{N_{\st},n_{\st}}(\bu_{\st}|\bv_{\st})\overline {K}^{(1/\mu)}_{N_{\so},n_{\so}}(\bu_{\so}|\bv_{\so}).
\end{multline}
Here the sum is taken over all possible partitions $\{\bv_{\so},\bv_{\st}\}\vdash\bv$ and $\{\bu_{\so},\bu_{\st}\}\vdash\bu$ without restrictions on the cardinalities of the subsets $N_{\so,\st}=\# \bu_{\so,\st}$ and  $n_{\so,\st}=\# \bv_{\so,\st}$.

Taking the limit $\kp\to\infty$ in \eqref{SP-prev} corresponds to the limit $\mu\to 1$. Thus, we obtain
\begin{multline}\label{SP-5}
S^{n,N}(\bv,\bu)=\mu^{N}\left(\frac{\rho_1}{\kappa^-}\right)^{N-n}\sum_{\substack{\{\bu_{\so},\bu_{\st}\}\vdash\bu\\ \{\bv_{\so},\bv_{\st}\}\vdash\bv }}
 \left(\frac{\rho_1}{\rho_2}\right)^{n_{\st}-N_{\st}}
\lambda_2(\bv_{\so})\lambda_2(\bu_{\st})\lambda_1(\bv_{\st})\lambda_1(\bu_{\so})
\\
\times f(\bv_{\so},\bv_{\st})f(\bu_{\st},\bu_{\so})
K^{(1)}_{N_{\st},n_{\st}}(\bu_{\st}|\bv_{\st})\overline {K}^{(1)}_{N_{\so},n_{\so}}(\bu_{\so}|\bv_{\so}).
\end{multline}
The sum is taken over partitions described above. Formally, we still have no restrictions on the cardinalities of the subsets. However, at least on MID in \eqref{SP-5} vanishes if $N_{\st}<n_{\st}$ or $N_{\so}<n_{\so}$ (see \eqref{defKdef1}, \eqref{CdefKdef2}).


\subsection{Scalar product with on-shell Bethe vector}

Up to now both vectors were off-shell. Let $\langle\Phi^{(1)}(\bv)|$ be on-shell Bethe vector, that means that the set $\bv$ satisfies Bethe equations \eqref{BEt}.
Now we can substitute the Bethe equations in the form
\be{BE2}
\lambda_1(\bv_{\st})f(\bv_{\so},\bv_{\st})=\alpha^{n_{\st}}\lambda_2(\bv_{\so})f(\bv_{\st},\bv_{\so}).
\ee
Then we obtain
\begin{equation}\label{SP-6}
S^{n,N}(\bv,\bu)=\mu^N\left(\frac{\rho_1}{\kappa^-}\right)^{N-n}\lambda_2(\bv)\sum_{\{\bu_{\so},\bu_{\st}\}\vdash\bu}\left(\frac{\rho_2}{\rho_1}\right)^{N_{\st}}
\lambda_2(\bu_{\st})\lambda_1(\bu_{\so})f(\bu_{\st},\bu_{\so})G(\bu_{\so},\bu_{\st}),
\end{equation}
where
\begin{equation}\label{Gp}
G(\bu_{\so},\bu_{\st})=\sum_{\{\bv_{\so},\bv_{\st}\}\vdash\bv }
\left(\frac{\alpha\rho_1}{\rho_2}\right)^{n_{\st}}f(\bv_{\st},\bv_{\so})K^{(1)}_{N_{\st},n_{\st}}(\bu_{\st}|\bv_{\st})\overline {K}^{(1)}_{N_{\so},n_{\so}}(\bu_{\so}|\bv_{\so}).
\end{equation}

Let us impose an additional constraint on the twist parameters: $\alpha=-\rho_2/\rho_1$. Then the sum over partitions in \eqref{Gp} can be explicitly computed via
\eqref{ML-3}:
\begin{equation}\label{Gp-res}
G(\bu_{\so},\bu_{\st})\Bigr|_{\alpha=-\rho_2/\rho_1}=(-1)^nf(\bv,\bu_{\so})K^{(1)}_{N,n}(\{\bu_{\so}-c,\bu_{\st}\}|\bv) .
\end{equation}
Respectively, the scalar product takes the form
\begin{multline}\label{SP-7}
S^{n,N}(\bv,\bu)=(-1)^n\mu^N\left(\frac{\rho_1}{\kappa^-}\right)^{N-n}\lambda_2(\bv)\\
\times\sum_{\{\bu_{\so},\bu_{\st}\}\vdash\bu}
\left(\frac{\rho_2}{\rho_1}\right)^{N_{\st}}
\lambda_2(\bu_{\st})\lambda_1(\bu_{\so})f(\bu_{\st},\bu_{\so})f(\bv,\bu_{\so})K^{(1)}_{N,n}(\{\bu_{\so}-c,\bu_{\st}\}|\bv).
\end{multline}
The sum over partitions of the set $\bu$ in \eqref{SP-7} is computed via proposition~\ref{CorSum}.  For this, it is enough to use representation for
MID in the form \eqref{Kmodeat01}
\be{KNn1z}
K_{N,n}^{(1)}(\bar w|\bv)=\lim_{z\to 1}(1-z)^{n-N}\Delta(\bar w)\Delta'(\bar\eta)\det_N\left(\frac{f(w_j,\bv)}{g(w_j,\bar\eta_k)}-zh(w_j,\bar\eta_k)\right),
\ee
where $\bar w=\{\bu_{\so}-c,\bu_{\st}\}$ for any fixed partition $\{\bu_{\so},\bu_{\st}\}\vdash\bu$.
Then we satisfy the condition of proposition~\ref{CorSum}. Recall also that $\bet=\{\eta_1,\dots,\eta_N\}$ are arbitrary complex numbers.
 Thus, we find
\begin{equation}\label{SP-res}
S^{n,N}(\bv,\bu)= (-1)^n\mu^N\left(\frac{\kappa^-}{\rho_1}\right)^{n}\lambda_2(\bv)\Delta'(\bet)\Delta(\bu)
\lim_{z\to 1} (1-z)^{n-N} \det_N \mathcal{N}_{jk}(z),
\end{equation}
where
\begin{equation}\label{Njk}
\mathcal{N}_{jk}(z)
=(-1)^{N-1}\hat\lambda_1(u_j)\left(h(\bet_k,u_j)-z\frac{f(\bv,u_j)}{g(\bet_k,u_j)}   \right)
+\hat\lambda_2(u_j)\left(\frac{f(u_j,\bv)}{g(u_j,\bet_k)}-zh(u_j,\bet_k)\right),
\end{equation}
and
\be{hlam}
\hat\lambda_\ell(u)=\frac{\rho_\ell}{\kappa^-}\lambda_\ell(u), \qquad \ell=1,2.
\ee

\subsection{How to take the limit $z\to 1$}

To take the limit $z\to 1$ in equation \eqref{SP-res}, we set $\eta_k=v_k$ for $k=1,\dots,n$ in the matrix \eqref{Njk}. Then
\begin{equation}\label{SP-res1}
S^{n,N}(\bv,\bu)=(-1)^n\mu^N\left(\frac{\kappa^-}{\rho_1}\right)^{n}\lambda_2(\bv)\Delta'(\bv)\Delta'(\bet)g(\bv,\bet)\Delta(\bu)
\lim_{z\to 1} (1-z)^{n-N} \det_N \mathcal{N}_{jk}(z),
\end{equation}
where now and further $\bet=\{\eta_{n+1},\dots,\eta_N\}$. The matrix $\mathcal{N}_{jk}(z)$ now consists of two parts:
\be{Njk-two}
\begin{aligned}
&\mathcal{N}_{jk}(z)=\mathcal{N}_{jk}^{(1)}(z),\qquad k=1,\dots,n,\\
&\mathcal{N}_{jk}(z)=\mathcal{N}_{jk}^{(2)}(z),\qquad k=n+1,\dots,N.
\end{aligned}
\ee
Here
\begin{multline}\label{Njk1}
\mathcal{N}_{jk}^{(1)}(z)=(-1)^{N-1}\hat\lambda_1(u_j)h(\bv,u_j)\left( \frac{h(\bet,u_j)}{h(v_k,u_j)}-z \frac{g(v_k,u_j)}{g(\bet,u_j)}   \right)\\
+\hat\lambda_2(u_j)h(u_j,\bv)\left(\frac{g(u_j,v_k)}{g(u_j,\bet)}-z\frac{h(u_j,\bet)}{h(u_j,v_k)}\right),\qquad k=1,\dots,n,
\end{multline}
and
\begin{multline}\label{Njk222}
\mathcal{N}_{jk}^{(2)}(z)=(-1)^{N}\hat\lambda_1(u_j)h(\bv,u_j)\left( \frac{z}{g(\bet_k,u_j)} -h(\bet_k,u_j)  \right)\\
+\hat\lambda_2(u_j)h(u_j,\bv)\left(\frac{1}{g(u_j,\bet_k)}-zh(u_j,\bet_k) \right),\qquad k=n+1,\dots,N.
\end{multline}
Now we can transform $\det_N \mathcal{N}_{jk}(z)$ using corollary~\ref{01-Pr-2} (see, in particular, \eqref{01-dAdA-1a}).  Then
\be{NtN}
 \det_N  \mathcal{N}(z)= (z-1)^{N-n}\det_N \widetilde{\mathcal{N}}(z).
\ee
Here $\widetilde{\mathcal{N}}_{jk}(z)=\mathcal{N}^{(1)}_{jk}(z)$ for $k=1,\dots,n$, and
\be{newN}
\widetilde{\mathcal{N}}_{jk}(z)= (-1)^{n+1}\frac{\hat\lambda_1(u_j)h(\bv,u_j)}{g(u_j,\bar\eta_k)}-\hat\lambda_2(u_j)h(u_j,\bv)h(u_j,\bar\eta_k),
\qquad k=n+1,\dots,N.
\ee
Substituting this result into \eqref{SP-res1} we obtain
\begin{equation}\label{SP-res2}
S^{n,N}(\bv,\bu)=(-\mu)^N\left(\frac{\kappa^-}{\rho_1}\right)^{n}\lambda_2(\bv)\Delta'(\bv)\Delta'(\bet)g(\bv,\bet)\Delta(\bu)
 \det_N \widetilde{\mathcal{N}}_{jk}(1).
\end{equation}
In this representation, the parameters $\bet=\{\eta_{n+1},\dots,\eta_N\}$ remain arbitrary complex numbers.

\subsection{Particular case\label{SS-PC}}

We did not impose any restrictions on the parameters $\bu$ in the representation \eqref{SP-res2}. Therefore, the vector $|\Phi^{(2)}(\bu)\rangle$, generally speaking, is off-shell. Consider a particular case $\alpha=1$. Then the dual vector $\langle\Phi^{(1)}(\bv)|$ is an eigenvector of the ordinary transfer matrix $t_{11}(z)+t_{22}(z)$. On the other hand, we find $\rho_1=-\rho_2$ from the condition $\alpha=-\rho_2/\rho_1$.
Then the inhomogeneous term vanishes in the modified Bethe equations \eqref{MBE}. We obtain
\be{BEred}
\hat\lambda_1(u_j)= (-1)^N\frac{\kappa+\rho}{\kb-\rho}\;\frac{h(u_j,\bu)}{h(\bu,u_j)}\,\hat\lambda_2(u_j),\qquad j=1,\dots,N,
\ee
where $\rho=\rho_1=-\rho_2$. At the same time,  the Bethe vectors $|\Phi^{(2)}(\bu)\rangle$ remain modified in this particular case. They are still given by \eqref{Phinu}.

We now require the vector $|\Phi^{(2)}(\bu)\rangle$ to be an on-shell modified Bethe vector. For this, it is sufficient to substitute conditions \eqref{BEred} into representation \eqref{SP-res2}. After simple algebra, we obtain
\begin{equation}\label{SP-res3}
S^{n,N}(\bv,\bu)=(-\mu)^N\left(\frac{\rho}{\kappa^-}\right)^{N-n}\lambda_2(\bv)\lambda_2(\bu)h(\bu,\bv)\Delta'(\bv)\Delta'(\bet)g(\bv,\bet)\Delta(\bu)
 \det_N \Omega_{jk}.
\end{equation}
Here the matrix $\Omega$ consists of two parts:
\begin{equation}\label{Omjk1}
\Omega_{jk}=\frac{g(u_j,v_k)}{g(u_j,\bet)}-\frac{h(u_j,\bet)}{h(u_j,v_k)}
-V_j\left( \frac{h(\bet,u_j)}{h(v_k,u_j)}- \frac{g(v_k,u_j)}{g(\bet,u_j)}   \right)
,\qquad k=1,\dots,n,
\end{equation}
and
\begin{equation}\label{Omjk2}
\Omega_{jk}=(-1)^{N+n+1}\frac{V_j}{g(u_j,\bar\eta_k)}-h(u_j,\bar\eta_k),
\qquad k=n+1,\dots,N,
\end{equation}
where
\be{Vj}
V_j=\frac{\kappa+\rho}{\kb-\rho}\;\frac{h(u_j,\bu)h(\bv,u_j)}{h(\bu,u_j)h(u_j,\bv)}.
\ee
It is easy to see that in a particular case $n=N$, we reproduce a determinant representation for the scalar product of the usual on-shell and the twisted on-shell Bethe vectors
\cite{Sla89}.


\section*{Conclusion}

We have considered overlaps of Bethe vectors with a diagonal twist and modified Bethe vectors. We have shown that under one additional condition on the twist parameters, such an overlap has a determinant representation. At the same time, the modified Bethe vector can remain an off-shell vector. In this sense, the result obtained is analogous to the scalar product of on-shell and off-shell Bethe vectors \cite{Sla89}.

However, despite this similarity, the resulting determinant representation has an entirely new structure. Namely, it contains a set of arbitrary parameters. This representation was obtained thanks to new formulas for the MID. Depending on the specific task, these parameters can be chosen in the most convenient way. Such formulas can find applications in models with higher symmetry rank \cite{GroLMR21}.

The resulting determinant representation for the overlap remains valid if the modified Bethe vector is the eigenvector of the modified transfer matrix. In this case, the parameters of the modified vector satisfy the inhomogeneous Bethe equations \eqref{MBE}.  We have considered only one specific case, when the inhomogeneous term in the Bethe equations disappears. The more general case requires further study.

One of the apparent directions for further research is the application of the results obtained to nonequilibrium physics. For this, it is necessary to consider the overlaps in the thermodynamic limit and calculate the density of states. Another area to consider is overlaps under diagonal and off-diagonal boundary conditions in the models on a segment. It is natural to expect that, in this case, the results are expressed in terms of the modified Tsuchiya determinant \cite{Tsu98}. We hope that these formulas are easy to obtain using the method of reducing overlaps to a system of linear equations \cite{BElS19b}.

\section*{Acknowledgements}
The work was performed at the Steklov Mathematical Institute of Russian Academy of Sciences,
Moscow. This work is supported by the Russian Science Foundation under grant 19-11-00062.


\appendix

\section{Properties of the MID}\label{A-PMID}

\subsection{Transformations of the MID}\label{SS-TMID}

In this appendix, we list several  properties of the MID. A more detailed list together with the proofs can be found in \cite{BelSV18sc}.

The MID and the conjugated one are related by
\be{c-c}
\overline{K}_{n,m}^{(z)}(\bu|\bv)=(1-z)^{m-n}K_{m,n}^{(z)}(\bv|\bu).
\ee

The MID possesses the following property under the shift of one of the set of parameters:
\be{Kinv1}
K_{n,m}^{(z)}(\bu-c|\bv)=\frac{(-z)^{n}(1-z)^{m-n} }{f(\bv,\bu)}K_{m,n}^{(1/z)}(\bv|\bu).
\ee
\be{cKinv1}
\overline{K}_{n,m}^{(z)}(\bu+c|\bv)=\frac{(-z)^{n}(1-z)^{m-n} }{f(\bu,\bv)}\overline{K}_{m,n}^{(1/z)}(\bv|\bu).
\ee

Some bilinear combinations of the MID reduce to a new MID.

\begin{prop}\label{GenML}
Let $\bxi$, $\bu$, and $\bv$ be sets of arbitrary complex numbers such that $\#\bxi=l$, $\#\bu=n$, and $\#\bv=m$.
Then
\be{ML-1}
\sum_{\{\bxi_{\so},\bxi_{\st}\}\vdash\bxi}z_2^{l_{\so}}
K_{n,l_{\so}}^{(z_1)}(\bu|\bxi_{\so})K_{m,l_{\st}}^{(z_2)}(\bv|\bxi_{\st})f(\bxi_{\st},\bxi_{\so})f(\bu,\bxi_{\st})
=K_{n+m,l}^{(z_1z_2)}(\{\bu,\bv\}|\bxi).
\ee
Here  $l_{\so}=\#\bxi_{\so}$ and $l_{\st}=\#\bxi_{\st}$. The sum is taken with respect to all partitions $\{\bxi_{\so},\bxi_{\st}\}\vdash\bxi$. There is no any restriction on the cardinalities of the subsets.
\end{prop}

Replacing $K_{n,l_{\so}}^{(z_1)}(\bu|\bxi_{\so})$ in \eqref{ML-1} by the conjugated MID via \eqref{c-c} and \eqref{Kinv1} we obtain
\be{ML-2}
\sum_{\{\bxi_{\so},\bxi_{\st}\}\vdash\bxi}\left(-\frac{z_2}{z_1}\right)^{l_{\so}}
\overline{K}_{n,l_{\so}}^{(z_1)}(\bu|\bxi_{\so})K_{m,l_{\st}}^{(z_2)}(\bv|\bxi_{\st})f(\bxi_{\st},\bxi_{\so})
=f(\bxi,\bu)K_{n+m,l}^{(z_1z_2)}(\{\bu-c,\bv\}|\bxi).
\ee
Setting here $z_1=z_2=1$ we arrive at
\be{ML-3}
\sum_{\{\bxi_{\so},\bxi_{\st}\}\vdash\bxi}(-1)^{l_{\so}}
\overline{K}_{n,l_{\so}}^{(1)}(\bu|\bxi_{\so})K_{m,l_{\st}}^{(1)}(\bv|\bxi_{\st})f(\bxi_{\st},\bxi_{\so})
=f(\bxi,\bu)K_{n+m,l}^{(1)}(\{\bu-c,\bv\}|\bxi).
\ee

\subsection{Other determinants related to the MID}\label{SS-ODRMID}

\begin{prop}\label{01-Pr-1}
Let $\bu=\{u_1,\dots,u_N\}$, $\bar\eta=\{\eta_1,\dots,\eta_N\}$, and $z$ be arbitrary complex numbers. Let $F_k^{(1)}(u)$ and $F_k^{(2)}(u)$, $k=1,\dots,N$, be two sets of functions
\be{01-B1}
F_k^{(1)}(u)=\phi_1(u)\left(\frac z{g(\bar\eta_k,u)}-h(\bar\eta_k,u)\right)+\phi_2(u)\left(\frac1{g(u,\bar\eta_k)}-z h(u,\bar\eta_k)\right),
\ee
\be{01-B2}
F_k^{(2)}(u)=\frac{(-1)^{N-1}\phi_1(u)}{g(u,\bar\eta_k)}-\phi_2(u)h(u,\bar\eta_k).
\ee
Here $\phi_\ell(z)$ ($\ell=1,2$) are two arbitrary functions. Let us compose two $N\times N$ matrices $\hat F^{(1)}$ and $\hat F^{(2)}$ with the entries
\be{twomat}
\hat F_{jk}^{(1)}=F_k^{(1)}(u_j),\qquad \hat F_{jk}^{(2)}=F_k^{(2)}(u_j).
\ee
 Then
\be{01-dAdA}
\det_N \hat F^{(1)} = (z-1)^N\det_N \hat F^{(2)}.
\ee
\end{prop}

\textsl{Proof.}  Obviously, both determinants can be presented in the form
\be{01-form}
\det_N \hat F^{(\ell)}=\sum_{\{\bu_{\so},\bu_{\st}\}\vdash \bu}X^{(\ell)}(\bu_{\so},\bu_{\st})\phi_1(\bu_{\so})\phi_2(\bu_{\st}),\qquad \ell=1,2,
\ee
where the coefficients $X^{(\ell)}(\bu_{\so},\bu_{\st})$ do not depend on $\phi_1$ and $\phi_2$. In this equation, we used the shorthand notation for the products
of the functions $\phi_\ell(u)$, $\ell=1,2$. Since $\phi_\ell$ are arbitrary functions, equation \eqref{01-dAdA}
holds if and only if
\be{01-XX}
X^{(1)}(\bu_{\so},\bu_{\st})=(z-1)^N X^{(2)}(\bu_{\so},\bu_{\st})
\ee
for an arbitrary partition $\{\bu_{\so},\bu_{\st}\}\vdash \bu$. However, it is easy to see that
without loss of generality, it is enough to prove \eqref{01-XX} for $\bu_{\so}=\{u_1,\dots,u_p\}$ and $\bu_{\st}=\{u_{p+1},\dots,u_N\}$. Here $p$ is an arbitrary integer from the set $\{0,1,\dots,N\}$.

To obtain the coefficients $X^{(\ell)}(\bu_{\so},\bu_{\st})$,  we should set $\phi_2(u_j)=0$ for $j=1,\dots,p$ and $\phi_1(u_j)=0$ for $j=p+1,\dots,N$ in \eqref{01-B1} and \eqref{01-B2}. We obtain
\be{matrices}
X^{(\ell)}(\bu_{\so},\bu_{\st})=\det_N \Phi^{(\ell)}, \qquad \ell=1,2,
\ee
where
\be{01-X1}
\begin{aligned}
&\Phi^{(1)}_{jk}=\frac z{g(\bar\eta_k,u_j)}-h(\bar\eta_k,u_j), \qquad j=1,\dots,p,\\
&\Phi^{(1)}_{jk}=\frac1{g(u_j,\bar\eta_k)}-z h(u_j,\bar\eta_k), \qquad j=p+1,\dots,N,
\end{aligned}
\ee
and
\be{01-X2}
\begin{aligned}
&\Phi^{(2)}_{jk}=\frac{(-1)^{N-1}}{g(u_j,\bar\eta_k)},\qquad j=1,\dots,p,\\
&\Phi^{(2)}_{jk}=-h(u_j,\bar\eta_k), \qquad j=p+1,\dots,N.
\end{aligned}
\ee

Let $u_j=u'_j+c$ for $j=1,\dots,p$, and $u_j=u'_j$ for $j=p+1,\dots,N$. Using $1/g(u_j,\bar\eta_k)=h(u'_j,\bar\eta_k)$ and  $h(\bar\eta_k,u_j)=(-1)^{N-1}/g(u'_j,\bar\eta_k)$ (see \eqref{gfh-prop}) for $k=1,\dots,p$, we obtain
\be{matrices2}
X^{(\ell)}(\bu_{\so},\bu_{\st})=(-1)^{(\ell+p-1)N}\det_N \widetilde{\Phi}^{(\ell)}, \qquad \ell=1,2,
\ee
where
\be{01-wtPhi12}
\begin{aligned}
&\widetilde{\Phi}^{(1)}_{jk}=\frac1{g(u'_j,\bar\eta_k)}-z h(u'_j,\bar\eta_k),\\
&\widetilde{\Phi}^{(2)}_{jk}=h(u'_j,\bar\eta_k),
\end{aligned}
\qquad\qquad j,k=1,\dots,N.
\ee
It is easy to see that $\det_N\widetilde{\Phi}^{(2)}$ reduces to the Cauchy determinant:
\begin{equation}\label{01-X2-1}
\det_N\widetilde{\Phi}^{(2)}
=h(\bu',\bar\eta)\det_N\left(\frac1{h(u'_j,\eta_k)}\right)
= \frac1{\Delta(\bar\eta)\Delta'(\bu')}.
\end{equation}

The determinant of $\widetilde{\Phi}^{(1)}$ is computed in corollary~\ref{cor-ize}. Due to \eqref{Kmodeat1}, we have
\be{01-X1-4}
\det_N\widetilde{\Phi}^{(1)}=\frac{(1-z)^N}{\Delta(\bar\eta)\Delta'(\bu')}.
\ee
Comparing this equation with \eqref{01-X2-1} and using \eqref{matrices2} we see that
\be{01-XX-11}
X^{(1)}(\bu_{\so},\bu_{\st})=(z-1)^NX^{(2)}(\bu_{\so},\bu_{\st}).
\ee
\qed

\begin{cor}\label{01-Pr-2}
Let $0\le n\le N$, and let $\bu=\{u_1,\dots,u_N\}$, $\bar\eta=\{\eta_{n+1},\dots,\eta_N\}$, and $z$ be arbitrary complex numbers ($\bar\eta=\emptyset$ for $n=N$).
Let us compose two $N\times N$ matrices $\hat F^{(01)}$ and $\hat F^{(02)}$ with the entries
\be{twomat-01}
\hat F_{jk}^{(01)}=\begin{cases}
F_k^{(0)}(u_j),\qquad k=1,\dots,n,\\
F_k^{(1)}(u_j), \qquad k=n+1,\dots,N,
\end{cases}
\ee
\be{twomat-02}
\hat F_{jk}^{(02)}=\begin{cases}
F_k^{(0)}(u_j),\qquad k=1,\dots,n,\\
F_k^{(2)}(u_j), \qquad k=n+1,\dots,N.
\end{cases}
\ee
Here $F_k^{(1)}(u)$ and $F_k^{(2)}(u)$ respectively are given by \eqref{01-B1} and \eqref{01-B2}, while $F_k^{(0)}(u)$, $k=1,\dots,n$, are arbitrary functions. Then
\be{01-dAdA-1}
\det_N \hat F^{(01)} = (z-1)^{N-n}\det_N \hat F^{(02)}.
\ee
\end{cor}

\textsl{Proof.}  Developing the determinant $\det \hat F^{(01)}$  with respect to the first $n$ columns we obtain
\be{devel-det1}
\det_N \hat F^{(01)}=\sum_{\{\bu_{\so},\bu_{\st}\}\vdash\bu} (-1)^{P_{\so,\st}}\det_n \Big(F_k^{(0)}(u^{\so}_j)\Big) \det_{N-n} \Big(F_k^{(1)}(u^{\st}_j)\Big).
\ee
Here the sum is taken over partitions $\{\bu_{\so},\bu_{\st}\}\vdash\bu$ such that $\#\bu_{\so}=n$ and $\#\bu_{\st}=N-n$. Notation $u^{\so}_j$ (resp. $u^{\st}_j$) means
the $j$th element of the subset $\bu_{\so}$ (resp. $\bu_{\st}$). Finally, $P_{\so,\st}$ is the parity of the partition $\{\bu_{\so},\bu_{\st}\}\vdash\bu$.

Due to proposition~\ref{01-Pr-1}
\be{de-de}
\det_{N-n} \Big(F_k^{(1)}(u^{\st}_j)\Big)=(z-1)^{N-n}\det_{N-n} \Big(F_k^{(2)}(u^{\st}_j)\Big),
\ee
for arbitrary subset $\bu_{\st}$. Hence,
\be{devel-det}
\det_N \hat F^{(01)}=(z-1)^{N-n}\sum_{\{\bu_{\so},\bu_{\st}\}\vdash\bu} (-1)^{P_{\so,\st}}\det_n \Big(F_k^{(0)}(u^{\so}_j)\Big) \det_{N-n} \Big(F_k^{(2)}(u^{\st}_j)\Big).
\ee
Taking the sum over partitions we arrive at \eqref{01-dAdA-1}.\qed

In particular, setting
\be{phi12}
\phi_1(u)=(-1)^{N}\hat\lambda_1(u)h(\bv,u), \qquad  \phi_2(u)=\hat\lambda_2(u)h(u,\bv),
\ee
we obtain
\be{FN}
F_k^{(1)}(u_j)=\mathcal{N}^{(2)}_{jk}, \qquad k=n+1,\dots,N,
\ee
where $\mathcal{N}^{(2)}_{jk}$ is given by \eqref{Njk222}. Due to corollary~\ref{01-Pr-2}, we obtain
\be{01-dAdA-1a}
\det_N \hat F^{(01)} = (z-1)^{N-n}\det_N \hat F^{(02)}.
\ee
Here
\be{twomat-01a}
\hat F_{jk}^{(01)}=\begin{cases}
F_k^{(0)}(u_j),\qquad k=1,\dots,n,\\
\mathcal{N}^{(2)}_{jk}, \qquad k=n+1,\dots,N,
\end{cases}
\ee
\be{twomat-02a}
\hat F_{jk}^{(02)}=\begin{cases}
F_k^{(0)}(u_j),\qquad k=1,\dots,n,\\
(-1)^{n+1}\hat\lambda_1(u_j)\frac{h(\bv,u_j)}{g(u_j,\bar\eta_k)}-\hat\lambda_2(u_j)h(u_j,\bv)h(u_j,\bar\eta_k), \qquad k=n+1,\dots,N.
\end{cases}
\ee
and $F_k^{(0)}(u)$, $k=1,\dots,n$, are arbitrary functions.

\subsection{Summation formula}

Let a function $H(\bu)$ of $N$ variables $\bu=\{u_1,\dots,u_N\}$ be defined as follows:
\be{HPhi}
H(\bu)=\Delta(\bu)\det_N\Phi_k(u_j).
\ee
Here $\Phi_k(u)$ is a set of one-variable functions.

\begin{prop}\label{CorSum}
Let $\phi_1(u)$ and $\phi_2(u)$ be one-variable functions. Then
\be{Sum-partab}
\sum_{\{\bu_{\so},\bu_{\st}\}\vdash\bu}\phi_1(\bu_{\so})\phi_2(\bu_{\st})f(\bu_{\st},\bu_{\so})H(\{\bu_{\so}-c,\bu_{\st}\})
=\Delta(\bu)\det_N\Bigl(\phi_1(u_j)\Phi_k(u_j-c)+\phi_2(u_j)\Phi_k(u_j)\Bigr).
\ee
The sum is taken over all possible partitions $\{\bu_{\so},\bu_{\st}\}\vdash\bu$, and we used the shorthand notation for the products
of the functions $\phi_\ell(u)$, $\ell=1,2$.
\end{prop}

\textsl{Proof}. The proof is similar to the one of proposition~\ref{01-Pr-1}. The rhs of \eqref{Sum-partab} can be presented as follows:
\be{01-formA}
\Delta(\bu)\det_N\Bigl(\phi_1(u_j)\Phi_k(u_j-c)+\phi_2(u_j)\Phi_k(u_j)\Bigr)=\sum_{\{\bu_{\so},\bu_{\st}\}\vdash \bu}X(\bu_{\so},\bu_{\st})\phi_1(\bu_{\so})\phi_2(\bu_{\st}),
\ee
where the coefficients $X(\bu_{\so},\bu_{\st})$ do not depend on $\phi_1$ and $\phi_2$. Since $\phi_\ell$ are arbitrary functions, equation \eqref{Sum-partab}
holds if and only if
\be{01-XXA}
X(\bu_{\so},\bu_{\st})=f(\bu_{\st},\bu_{\so})H(\{\bu_{\so}-c,\bu_{\st}\}),
\ee
for an arbitrary partition $\{\bu_{\so},\bu_{\st}\}\vdash \bu$. Due to the symmetry of \eqref{Sum-partab} over $\bu$,
it is enough to prove \eqref{01-XXA} for $\bu_{\so}=\{u_1,\dots,u_p\}$ and $\bu_{\st}=\{u_{p+1},\dots,u_N\}$. Here $p$ is an arbitrary integer from the set $\{0,1,\dots,N\}$.

To obtain the coefficients $X(\bu_{\so},\bu_{\st})$,  we should set $\phi_2(u_j)=0$ for $j=1,\dots,p$ and $\phi_1(u_j)=0$ for $j=p+1,\dots,N$ in \eqref{Sum-partab}. We obtain
\be{matricesA}
X(\bu_{\so},\bu_{\st})=\Delta(\bu_{\so})\Delta(\bu_{\st})g(\bu_{\st},\bu_{\so})\det_N \widetilde{\Phi},
\ee
where
\be{01-X1A}
\begin{aligned}
&\widetilde{\Phi}_{jk}=\Phi_k(u_j-c), \qquad j=1,\dots,p,\\
&\widetilde{\Phi}_{jk}=\Phi_k(u_j), \qquad j=p+1,\dots,N.
\end{aligned}
\ee

Let $u_j=u'_j+c$ for $j=1,\dots,p$, and $u_j=u'_j$ for $j=p+1,\dots,N$. Using $g(x,y+c)=-1/h(y,x)$ for any $x$ and $y$, we obtain
\be{Delta}
\Delta(\bu_{\so})\Delta(\bu_{\st})g(\bu_{\st},\bu_{\so})=(-1)^{p(N-p)}\frac{\Delta(\bu'_{\so})\Delta(\bu'_{\st})}{h(\bu'_{\so},\bu'_{\st})}
=\frac{\Delta(\bu')}{f(\bu'_{\so},\bu'_{\st})}.
\ee
Then \eqref{matricesA} takes the form
\be{matricesAA}
X(\bu_{\so},\bu_{\st})=\frac{\Delta(\bu')}{f(\bu'_{\so},\bu'_{\st})}\det_N \Phi_k(u'_j)=\frac{H(\bu')}{f(\bu'_{\so},\bu'_{\st})}.
\ee
Turning back to the original variables and using $f(x-c,y)=1/f(y,x)$ for any $x$ and $y$ we arrive at \eqref{01-XXA}. \qed

We deal with a particular case of the sum \eqref{Sum-partab} in equation \eqref{SP-7}. Indeed, let
\be{abP}
\phi_1(u)= \lambda_1(u)f(\bv,u), \qquad \phi_2(u)=\frac{\rho_2}{\rho_1}\lambda_2(u), \qquad
\Phi_k(u)=\frac{f(u,\bv)}{g(u,\bet_k)}-z h(u,\bet_k),
\ee
where $\bet=\{\eta_1,\dots,\eta_N\}$ are arbitrary complex numbers. Then
\be{hu-sp}
\Delta'(\bet) H(\bu)=(1-z)^{N-n}K^{(z)}_{N,n}(\bu|\bv).
\ee
Hence, due to proposition~\ref{CorSum}, we obtain
\begin{multline}\label{PC-1}
\sum_{\{\bu_{\so},\bu_{\st}\}\vdash\bu}
\left(\frac{\rho_2}{\rho_1}\right)^{N_{\st}}
\lambda_2(\bu_{\st})\lambda_1(\bu_{\so})f(\bu_{\st},\bu_{\so})f(\bv,\bu_{\so})K^{(1)}_{N,n}(\{\bu_{\so}-c,\bu_{\st}\}|\bv)\\
=(1-z)^{N-n}\Delta'(\bet)\Delta(\bu)\det_N \mathcal{M}_{jk},
\end{multline}
where
\begin{multline}\label{Mjk}
\mathcal{M}_{jk}=\lambda_1(u_j)f(\bv,u_j)\left(\frac{f(u_j-c,\bv)}{g(u_j-c,\bet_k)}-z h(u_j-c,\bet_k)\right)\\
+\frac{\rho_2}{\rho_1}\lambda_2(u_j)\left(\frac{f(u_j,\bv)}{g(u_j,\bet_k)}-z h(u_j,\bet_k)\right).
\end{multline}
Using equations \eqref{gfh-prop} we easily transform this result to representation \eqref{SP-res}.


\begin{thebibliography}{99}
%
\bibitem{FadST79} L.D. Faddeev, E.K. Sklyanin and L.A. Takhtajan, \textsl{Quantum Inverse Problem. I},
 Theor. Math. Phys. {\bf 40} (1979) 688--706.
%
\bibitem{FadLH96} L.D. Faddeev, \textit{How Algebraic Bethe Ansatz works
for integrable model}, in: Les Houches Lectures \textsl{Quantum Symmetries}, eds A. Connes
et al, North Holland, (1998) 149, \texttt{arXiv:hep-th/9605187}.
%
\bibitem{BogIK93L}V.E. Korepin, N.M. Bogoliubov,
A.G. Izergin, {\sl Quantum Inverse Scattering Method and Correlation Functions}, Cambridge: Cambridge Univ.
Press, 1993.
%
%
\bibitem{BelC13} S. Belliard and N. Cramp\'e, \textsl{Heisenberg XXX model with general boundaries: Eigenvectors from Algebraic Bethe ansatz},
SIGMA {\bf 9} (2013) 72, \texttt{arXiv:1309.6165}.
%
\bibitem{CYSW13a}  J. Cao, W. Yang , K. Shi  and Y. Wang,
 \textsl{Off-diagonal Bethe ansatz and exact solution a topological spin ring},
Phys. Rev. Lett. {\bf 111} (2013) 137201,  \texttt{arXiv:1305.7328}.
%
\bibitem{CYSW13c}  J. Cao, W. Yang , K. Shi  and Y. Wang,
 \textsl{Off-diagonal Bethe ansatz for exactly solvable models},
Springer, Berlin, Heidelberg 2015.
%
\bibitem{BelSV18sc} S. Belliard, N.A. Slavnov, B. Vallet, \textsl{Scalar product of twisted XXX modified Bethe vectors},
J. Stat. Mech., {\bf2018}:9 (2018) 93103, \texttt{arXiv:1805.11323}.
%
\bibitem{BelSV18} S. Belliard, N. A. Slavnov, B. Vallet, \textsl{Modified algebraic Bethe Ansatz: twisted XXX case}, SIGMA, {\bf14} (2018) 54, \texttt{arXiv:1804.00597}.
%
\bibitem{BelS19} S. Belliard, N.A. Slavnov, \textsl{Scalar products in twisted XXX spin chain. Determinant representation},
SIGMA {\bf15} (2019) 066, \texttt{arXiv:1906.06897}.
%
\bibitem{Gau72} M. Gaudin, \textsl{Mod\`eles exacts en m\'ecanique statistique: la m\'ethode de Bethe et ses g\'en\'eralisations}, Preprint, Centre d'Etudes Nucl\'eaires de Saclay, CEA-N-{\bf 1559}:1 (1972).
%
\bibitem{Gaud83} M. Gaudin, \textsl{La Fonction d'Onde de Bethe}, Paris: Masson, 1983.
%
\bibitem{Kor82} V.E. Korepin, \textsl{Calculation of norms of Bethe wave functions}, Comm. Math. Phys. {\bf 86} (1982) 391--418.
%
\bibitem{Sla89} N.A. Slavnov, \textsl{Calculation of scalar products of wave functions and form factors in the framework of the algebraic Bethe ansatz},
Theor. Math. Phys. {\bf 79} (1989) 502--508.
%
\bibitem{BelP15} S. Belliard, R.A. Pimenta, \textsl{Slavnov and Gaudin--Korepin Formulas for Models without U(1) Symmetry: the Twisted XXX Chain},
SIGMA {\bf11} (2015) 099,  \texttt{arXiv:1506.06550}.
%
\bibitem{GorZZ14} A. Gorsky, A. Zabrodin, A. Zotov, \textsl{Spectrum of quantum transfer matrices via classical many-body systems},
JHEP (2014)  070, \texttt{arXiv:1310.6958}.
%
\bibitem{Ize87} A. G. Izergin, \textsl{Partition function of the six-vertex model in a finite volume},
Sov. Phys. Dokl. {\bf 32} (1987) 878--879.

%
\bibitem{GroLMR21}
N. Gromov, F. Levkovich-Maslyuk, P. Ryan, \textsl{Determinant form of correlators in high rank integrable spin chains via separation of variables},
JHEP \textbf{05} (2021) 169, \texttt{arXiv:2011.08229}.
%
\bibitem{Tsu98} O. Tsuchiya, \textsl{Determinant formula for the six-vertex model with reflecting end}, J. Math. Phys. {\bf 39} (1998) 5946,
\texttt{arXiv:solv-int/9804010}.
%
%
\bibitem{BElS19b} S.~Belliard, N.A.~Slavnov, \textsl{Why scalar products in the algebraic Bethe ansatz have determinant representation}, JHEP,
{\bf 2019}:10 (2019) 103, \texttt{arXiv:1908.00032}.




%
%
%
%
%
%

%

\end{thebibliography}
\end{document}